# Comparison of the atomic level structure of the plastic crystalline and the liquid phases of $CBr_2Cl_2$: neutron diffraction and Reverse Monte Carlo modeling

**Szilvia Pothoczki[1], László Temleitner[1], Luis Carlos Pardo[2], Gabriel Julio Cuello[3], Muriel Rovira-Esteva[2] and Josep Lluis Tamarit[2]**

[1]Institute for Solid State Physics and Optics, Wigner Research Centre for Physics, Hungarian Academy of Sciences, PO Box 49, H-1525 Budapest, Hungary
[2]Grup de Caracteritzacio de Materials, Depertament de Fisica i Enginyeria Nuclear, ETSEIB, Universitat Politecnica de Catalunya, Diagonal 647, E-08028 Barcelona, Catalonia, Spain
[3]Institut Laue Langevin, BP 156, Rue Jules Horowitz 6, F-38042 Grenoble Cedex 9, France

E-mail: pothoczki.szilvia@wigner.mta.hu, temleitner.laszlo@wigner.mta.hu

**Abstract.** Neutron diffraction results obtained for plastic crystalline dichlorodibromomethane ($CBr_2Cl_2$) have been modelled by means of the Reverse Monte Carlo method. Comparison with its liquid phase is provided at several levels of the atomic structure (total scattering structure factors, partial radial distribution functions, orientational and dipole-dipole correlations). Results reveal that the relative orientation of neighbouring molecules largely depends on the steric effect. The small dipole moment has not as strong influence as the steric effect on the short range order. Our observations fit well with earlier findings presented for the series $CBr_nCl_{4-n}$ (n=0, 1, 2, 4).

1. Introduction

During the last decade the structures of disordered (liquid) or partially disordered (plastic crystalline) phases of the series of $CCl_4$ [1-4], $CBrCl_3$ [5-8], $CBr_2Cl_2$ [9-10], $CBr_4$ [11-14] halomethane compounds have been the object of many studies. One of the reasons for such an interest is that their molecular sizes are almost the same, with a nearly spherical molecular shape, whereas the molecular symmetry (and thus the dipole moment) covers a diverse range: $T_d$ ($CCl_4$, $CBr_4$), $C_{3v}$ ($CBrCl_3$) $C_{2v}$ ($CBr_2Cl_2$). This provides the possibility of a systematic investigation for revealing the role of molecular symmetry, as well as the steric effect, in shaping intermolecular correlations.

A common feature that makes these systems attractive is that an orientationally disordered crystalline phase – also called 'plastic crystal' – appears between the low-temperature ordered phase and the liquid phase [5,6,10]. In this special phase the molecular centres are located in a high-symmetry lattice (face centered cubic, body centered cubic or rhombohedral), but the molecules can rotate around their equilibrium position. The three characteristic phases (ordered, plastic crystalline and liquid) have been studied extensively for each halomethane mentioned above [1-16].

One of the few missing pieces of information is the detailed atomic level description of the plastic phase of $CBr_2Cl_2$. Here therefore we focus on this phase and provide comparison with the liquid phase of the material at every stage of structural analyses. Short range ordering is followed with an accentuated attention: we wish to go beyond partial radial distribution functions (prdf) and introduce special distance-dependent correlation functions that describe mutual arrangements of the molecules in both the plastic and the liquid phase of $CBr_2Cl_2$. Our calculations are based on structural models, containing thousands of atoms that are obtained from the Reverse Monte Carlo method [17-20]. Finally, we also aim to emphasize findings that may be connected with properties of plastic crystalline and liquid $CCl_4$, $CBr_4$ and $CBrCl_3$, thus completing the investigation on the $CBr_nCl_{4-n}$ (n=0, 1, 2, 4) family.

2. Experimental details

Sample of $CBr_2Cl_2$ was obtained from Sigma-Aldrich with purity of 99%. Neutron diffraction experiments were performed at Institute Laue-Langevin using the D4 [21] diffractometer with wavelength of 0.5022 Å. The temperature was set to 270 K, corresponding to the plastic crystalline face centered cubic phase. Empty cryostat, empty sample holder, vanadium rod and an absorbing sample were also measured to carry out corrections (detector efficiency, absorption, multiple scattering [22]) by means of CORRECT software [23]. The Bragg line profile parameters are determined by the measurement of nickel as standard.

Experimental details concerning the liquid phase are found in [9].

## 3. Details of RMC analysis

Modeling the disorder in crystalline systems can be performed by using various computer programs that apply the Reverse Monte Carlo algorithm designed for this purpose [24-26]. The primary aim of the present study is to reveal intermolecular correlations: this kind of information is contained mostly in the low Q diffuse scattering part of the data (approximately below 6 Å$^{-1}$, see, e.g., Ref. [27]). This is why comparison in the Q-space is suggested between experimental and simulated datasets. For this purpose, the RMCPOW [24] algorithm appears to be the most suitable, since it calculates the total scattering pattern by a 3 dimensional Fourier-transformation (FT) and separates Bragg and diffuse scattering contributions. In RMCPOW, the experimental Q-space information is calculated directly from the particle positions, using 'exp(-iQr)' sums; that is, the radial distribution function is not computed. This is the reason why RMCPOW does not require data over wide Q-ranges. The two contributions mentioned above are treated according to their nature: the Bragg-part intensities are convoluted by the instrumental resolution function; for the diffuse part each contributing point in the reciprocal space is averaged over the volume (for details, see [24]). In the sense of other aspects, it follows the standard RMC procedure [17]. Using this procedure, even relatively small super cells can be considered, without any convolution of the experimental data. On the other hand, if one wishes to extend the calculation for the entire measured momentum transfer range then enormous computational efforts are necessary, due to the 3D FT. A one dimensional Fourier-transformation is much faster, but, in general, we need to convolute the dataset (usually with a step function corresponding to the half of the box length) before the procedure.

However, this convolution is not necessary in the present case. It is clearly visible in fig. 1a, that the intensities of the Bragg-peaks are not so prominent in comparison with the diffuse-scattering contribution. This suggests that if we choose even a relatively small super cell, differences between experimental and 'convoluted experimental' datasets are comparable with the experimental uncertainties. This is why modeling may be performed for the original experimental data *without convolution*. If we choose the half box length to be 52.152 Å, the difference between convoluted and non-convoluted experimental datasets is of the order of the uncertainties over the whole *Q*-range (see fig. 1d). Noticeable differences can only be found between 1 and 2 Å$^{-1}$; note that even these are not larger than 1% of the measured intensity. That is, if we simulate a system of this size, which corresponds to a 12x12x12 multiplication of the fcc Bravais-cell (34560 atoms), then we can apply the original (non-convoluted) measured data.

The final RMC simulation for the plastic crystalline phase has therefore been carried out by the RMC_POT [20] software (this procedure uses 1D Fourier-transform). To accelerate convergence, we applied molecular moves (translation, rotation and small individual moves for each atom) instead of single moves. This not only accelerated, but kept the geometry of the molecule fixed and prevented the calculation from being stuck in.

Below we provide the detailed process for the plastic phase of $CBr_2Cl_2$ to reach the final configurations:

(1) The lattice constant of the plastic crystalline phase, arising from the result of indexing the Bragg profile, were initially set to 8.672 Å.

(2) A short RMCPOW simulation was performed, using a supercell of 4x4x4 times of the unit cell up to 4 Å$^{-1}$, in order to check the settings. The following set of distance constraints, so-called fixed neighbour constrains (fnc) [19] were applied to maintain the appropriate molecular structure: 1.87 Å < $d_{fnc}$(C-Br) < 2.04 Å; 1.69 Å < $d_{fnc}$(C-Cl) < 1.81 Å; 3.063 Å < $d_{fnc}$(Br-Br) < 3.303 Å; 2.85 Å < $d_{fnc}$(Br-Cl) < 3.09 Å; 2.79 Å < $d_{fnc}$(Cl-Cl) < 3.02 Å. Intermolecular closest approach distances (cutoffs) were set as $d_{cutoff}$(C-C) = 4.0 Å; $d_{cutoff}$(C-Cl) = 3.0 Å; $d_{cutoff}$(C-Br) = 3.0 Å; $d_{cutoff}$(Br-Br) = 3.0 Å; $d_{cutoff}$(Br-Cl) = 3.0 Å; $d_{cutoff}$(Cl-Cl) = 3.0 Å.

(3) During a series of 8x8x8 supercell RMCPOW simulations slightly different lattice parameters were used. Thus 8.692 Å resulted as the best fit (Bragg with diffuse part).

(4) In order to extend the simulation to the whole Q range (up to 20 Å$^{-1}$) as explained above, we used the RMC_POT software. The starting configuration was a 12x12x12 supercell containing 6912 molecules, namely 34560 atoms (ρ = 0.03046 atom/Å$^3$ corresponding to 52.152 Å half boxlength). We used molecular moves [19] (instead of atomic moves) to speed up the calculation. The cutoff's and fnc limits were the same as in (2).

(5) After reaching equilibrium (best goodness-of-fit while the ratio of accepted moves was still above 20% and $R_{wp}$=9.08%), independent configurations (separated by at least one successful move of each atom) were collected.

(6) Further calculations, for characterizing orientational ordering, were performed from 11 independent particle configurations.

Reverse Monte Carlo simulation for liquid $CBr_2Cl_2$ has been described in [9]; here we just intend to recall only some relevant details. In short, the initial model contained 2000 randomly oriented flexible molecules in a cubic box with periodic boundary conditions, using the appropriate atomic number density (0.03007 Å$^{-3}$) and molecular geometry. Molecules of distorted tetrahedral shape were held together also by means of 'fixed neighbours constraints'; the 'fnc' values were essentially the same as mentioned above for the plastic phase. A significant difference is that during the RMC_POT procedure [20] for the liquid, atomic moves were applied. For further calculations 25 independent configurations were collected.

Results for $CCl_4$ (for comparison only) have been taken from our previous RMC study on that liquid [3].

## 4. Results in Q space

The total scattering diffraction patterns of the plastic and liquid phases hardly differ (fig. 1a and 1b); the most remarkable variation can be found below 5 Å$^{-1}$, where the pattern intensity increases slightly for the plastic phase. This is a typical feature of the plastic crystal, where only few Bragg-peaks of the high symmetry (here Fm-3m) crystal appear and their intensities are damped due to large thermal displacements (c.f. the similar pattern of plastic crystalline $CBr_4$ [28]). At larger Q-values, oscillations related to the molecular structure remain. Taking into account the discussion of previous section, the present resolution is probably sufficient for capturing the main properties of the plastic crystal.

Figures 1a and 1c show the agreement between experimental and calculated powder diffraction patterns. The biggest differences appear at the 111 reflection, and a slight difference around 4 and 9 Å$^{-1}$. Concerning the small, but visible differences between experiment and RMCPOW, we can offer a couple of possible reasons for them: (1) the complicated molecular movement scheme introduced to RMCPOW may need even more computational time to explore configuration space to the necessary degree for achieving 'perfect' match; (2) there may be some small residual systematic errors in the corrected data. Note, however, that the level of agreement between measurement and model is at least comparable to that found in the literature for similar modeling studies [29,30].

Taking into account the time of measurement, this is most likely a small effect and does not bias our findings (see below).

At the beginning of the RMC calculations, all carbon atoms occupied the sites of an fcc lattice and the orientation of each molecule was the same. The faster (RMC_POT) algorithm does not distinguish between the Bragg- and diffuse-scattering contributions; in order to check if this is acceptable, we started an RMCPOW simulation for the final configuration. As it is obvious from fig. 1c, only a small discrepancy appears at about 1.2 Å$^{-1}$ in the diffuse scattering pattern.

Before moving to the next sections, it is worth considering the contribution of each partial pair correlation to the total neutron scattering structure factor. The weights of the prdf's are 2.8% for C-C, 11.6% for C-Br, 16.4% for C-Cl, 1.9% for Br-Br, 33.6% Br-Cl and 23.6% for Cl-Cl. These numbers indicate that the C-C partial can be determined mainly by taking into account 'steric' constraints during the simulations.

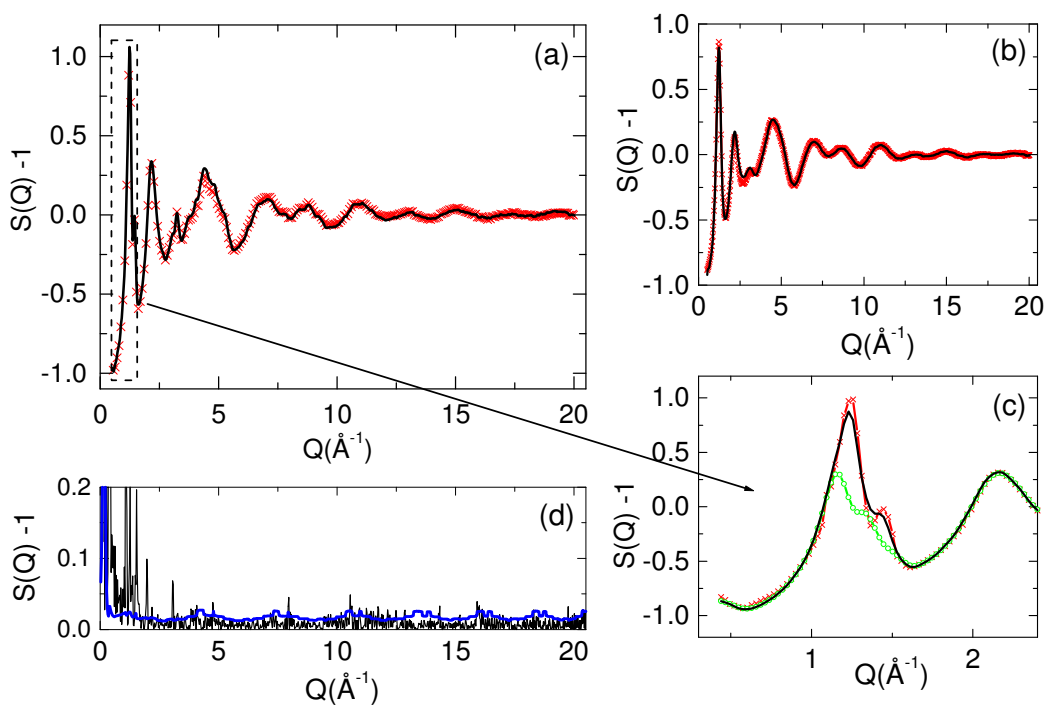

**Figure 1.** (a) Total scattering structure factors of plastic $CBr_2Cl_2$. (b) Total scattering structure factor of liquid $CBr_2Cl_2$ [9]. Red solid line with crosses: neutron diffraction data; black solid line: RMC. (c) Enlarged low Q part of total scattering structure factor of plastic $CBr_2Cl_2$, as calculated from the final configuration by RMCPOW. Red solid line with crosses: neutron diffraction data; black solid line: RMCPOW Bragg and diffuse; green line with open circles: diffuse part only. (d) Blue line: experimental uncertainty, black line: difference between the modulus of experimental and the convoluted (by the half of the box size) experimental datasets. (Intensities are multiplied by 5 for better visibility.)

5. Real-space analyses

5.1 Partial radial distribution functions

Partial radial distribution functions for the plastic and liquid phases of $CBr_2Cl_2$ are shown in figure 2. Recalling that nearly the same intramolecular parameters (fnc's) were used during the plastic and liquid phase RMC simulations, the two phases could not be different in this regard. As we saw in the previous section, the almost perfect high Q fit supports the chosen limits of 'fixed neighbour constraints'.

Concerning intermolecular correlations, the oscillations do not extend beyond 10-12 Å in the liquid phase, while they are more extended in the plastic phase; this is particularly valid for the centre-centre radial distribution function. Integrating up to the first minima of the C-C prdf's – 8.1 Å (plastic crystalline phase) and 8.5 Å (liquid phase) – coordination numbers of 13.28 (plastic crystalline phase) and 13.23 (liquid phase) were obtained. Furthermore the position of the first two peaks (around 6 and 11 Å) no significant differences can be noticed, suggesting that these two phases of $CBr_2Cl_2$ have a common short-range positional order.

The centre-halogen and the halogen-halogen radial distribution functions also show strong similarities between the liquid and the plastic phases. For all of these partials, two maxima emerge invariably between 3 and 9 Å, which correspond to the two halogens of the closest neighbouring molecules. Beyond this distance range the correlations disappear, the presence of longer range (beyond 20 Å) ordering has no any sign.

At this stage it is worth comparing our results with earlier findings on other halomethane compounds ($CCl_4$ [1, 3], $CBrCl_3$ [6], $CBr_4$ [13]). The largest difference occurs when the plastic phases of $CBr_2Cl_2$ and the $CBr_4$ are compared from the point of view of the C-Br prdf. In the case of $CBr_4$ this partial (not shown here; c.f middle panel of fig. 4 in [13]) behaves similarly to the monoclinic phase. Large oscillations extend up to 20 Å. In contrast, the two curves belonging to the plastic and the liquid phases run together for $CBr_2Cl_2$. Unfortunately, Caballero et al. [6] do not show this radial distribution function and therefore, we cannot provide a fully comprehensive comparison with the plastic phase of $CBrCl_3$ (comparison of the orientational correlations are supplied in next Section).

In addition, we plot the carbon-chlorine and chlorine-chlorine radial distribution functions of the liquid phase of $CCl_4$ [3]. Surprisingly, the carbon-chlorine peaks seem sharper for liquid $CCl_4$, but in terms of chlorine-chlorine correlations, liquid $CCl_4$ and plastic $CBr_2Cl_2$ are more similar. This resemblance confirms the earlier statement [1] that steric effects have a significant influence on the correlations in these molecule-based systems.

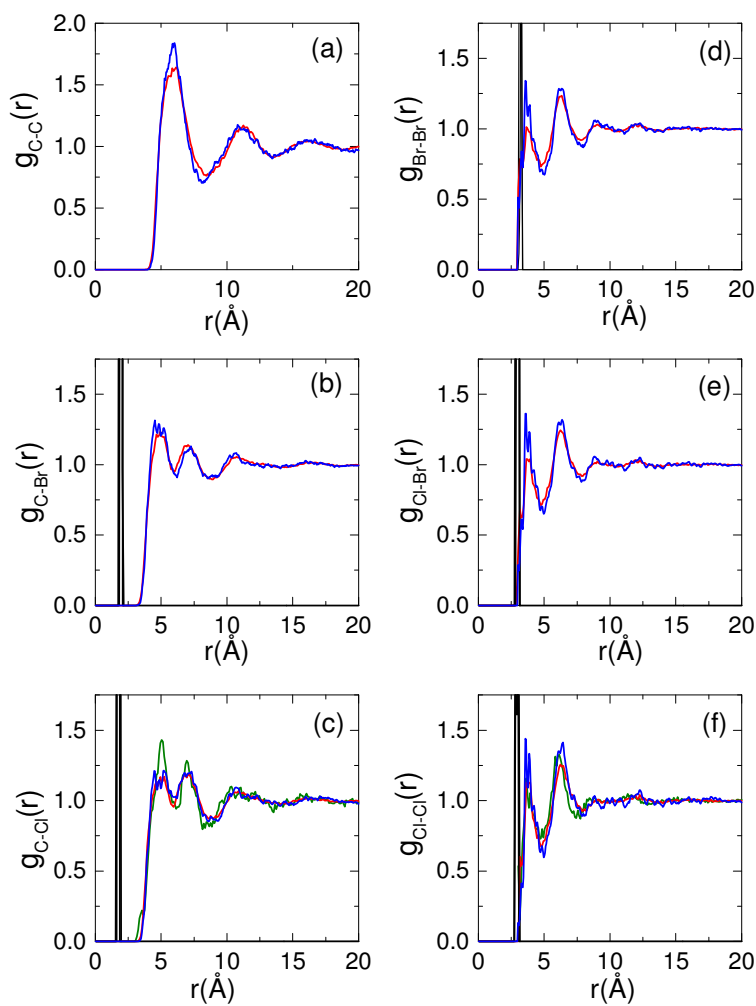

**Figure 2.** Partial radial distribution functions. (a) C-C; (b) C–Br; (c) C–Cl (d) Br-Br; (e) Br-Cl; (f) Cl-Cl. Solid black line: intramolecular part; solid blue line: intermolecular part for plastic crystalline $CBr_2Cl_2$; solid red line: intermolecular part for liquid $CBr_2Cl_2$; solid green line: intermolecular part for liquid $CCl_4$.

It is perhaps instructive to visualize the simulation box at this point. We produced a snapshot of the plastic phase where each atom is projected into one unit cell. For a better visibility of the fcc symmetry, carbon atoms were removed. Neither chlorine atoms nor bromine atoms composed continuous spots, as it is well visible in figure 3.a. We have to note that our initial configuration was a perfect crystal, that is, every molecule was oriented the same way (fcc ordered). The molecules markedly turned from their original direction. This process is largely facilitate by the 'molecular moves' option of the RMC_POT software [20].

Furthermore, we wish to draw attention to the sizes of the balls, which are related to the covalent radii of the atoms: this way, the significance of the excluded volume of the molecules is directly observable.

In figure 3.b a contour-plot of the atomic density of carbon atoms (in the {111} and {100} Miller-index planes) can be found. The fcc structure remains, but the extent of the spots indicates that carbon atoms diverge from their original positions, due to large thermal movements.

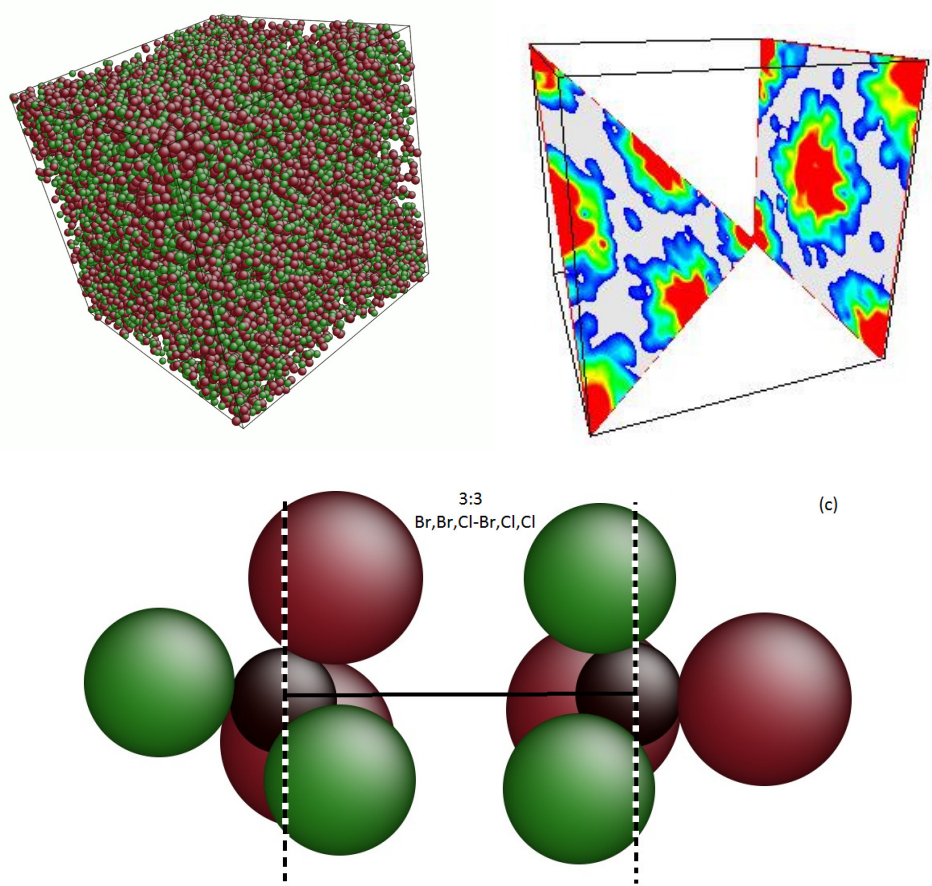

**Figure 3**. (a) Snapshot of the Bravais cell of the plastic phase from a simulated configuration. Red: bromine atoms; green: chlorine atoms. (b) Contour plot of the atomic density of carbon atoms. (Atomic configuration figures have been prepared by the ATOMEYE software [31]. Contour-plots have been made by the 'Confplot' software, which is part of the 'NFLP' program package.) (c) The most probable orientation of two neighbouring molecules in the plastic crystalline phase.

## 5.2 Orientational correlations

For the determination of molecular orientations between tetrahedral molecules we prefer the following classification, due to Rey [2]: given a pair of tetrahedral molecules we construct two parallel planes, each one containing the centres of these molecules. Every molecular pair was labelled into one of six orientational classes, depending on how many halogens from each molecule are found between these planes. This way, corner-to-corner (1:1); corner-to-edge (1:2); corner-to-face (1:3); edge-to-edge (2:2); edge-face (2:3); face-to-face (3:3) orientations may be distinguished. The clarity of this scheme permits its application to other systems consisting of tetrahedral molecules such as $CCl_4$ [3], $CBrCl_3$ [6] and $CBr_4$ [13]; it is also our intention here to compare our present results with these earlier works [3,6,13].

These various centre-centre distance-dependent correlation functions can be found in fig. 4 for plastic and liquid $CBr_2Cl_2$; only a slight difference can be noticed between the two phases. This statement was also true in the case of $CBrCl_3$ [6], supporting the strong resemblance between the two halomethane compounds. The pattern, in general, follows the one shown for $CBrCl_3$: the 2:2 orientations are the most populous beyond 8 Å, as was observed for the whole family of $CBr_nCl_{4-n}$ (n=0, 1, 2, 4). At short distances (up to 4.6 Å), i.e., well inside the first coordination shell (c.f. the centre-centre radial distribution functions in fig 2.a, where the first minimum is around 8.5 Å), the face-to-face (3:3) pairs dominate. It has to be pointed out that the actual number of 3:3 pairs is not so different at the shortest distances from what is found in the higher coordination shells, but the relative occurrence of such molecular pairs at those larger distances is much smaller. With increasing centre-centre distances the number of 3:3 molecular pairs sharply decreases and reaches the same level (below 10 %) as the 1:1 pairs.

The 1:2 and 2:3 groups markedly alternate. These oscillations are less sharp and shorter ranged than that we observed for $CCl_4$ (c.f. fig. 4b.). This behaviour is in accordance with the findings of Caballero et al. [6] for the plastic and liquid phases of $CBrCl_3$. These systems show analogous characteristics when considering either the plastic or the liquid phase; the extended, strong oscillations are typical only for $CCl_4$ (most probably, due to its nearly spherical contour). Note that in this respect, already the closest relative of $CCl_4$, $CBr_4$, is different.

In order to complete this analysis we mention that the occurrence of 1:3 (corner-to-face; a.k.a. "Apollo") orientation reaches almost 15% at around 6 Å in the plastic phase, otherwise it falls short of 10%. For the liquid phase they are even scarcer, staying below 10% nearly over the whole distance range.

In order to emphasize how the distributions diverge from their expected value, we normalized the intensities to their asymptotic values (observed at the highest carbon-carbon distances (see figure 4.c)). Thus each curve converges to unity. It is found that in the closest distance range (up to 5.3 Å) the 3:3 group is more than 25 times higher than its average value. Three other groups are twice more intense than their averages: the 2:3 group at 5.05 Å, the 1:3 group at 6.09 Å and the 1:1 group at 7.59 Å. This is in accordance with previous findings on the plastic phase of CCl$_4$ [1].

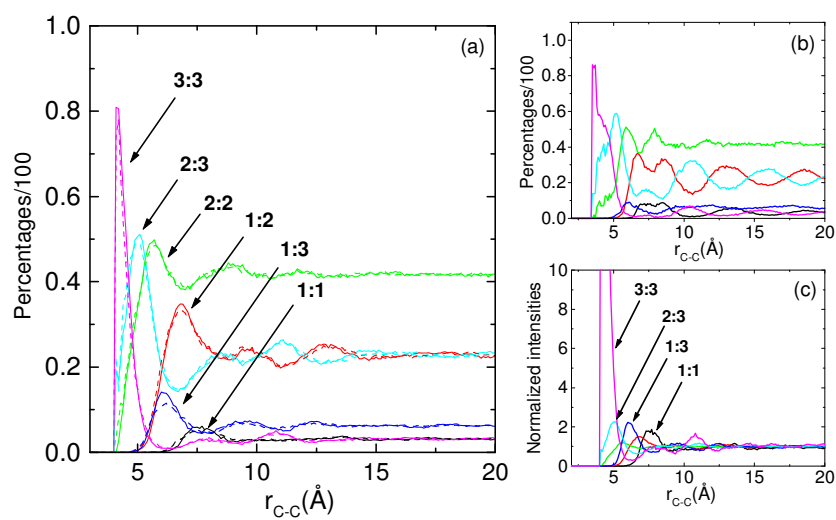

**Figure 4**. (a) Orientational correlation functions for plastic and liquid CBr$_2$Cl$_2$. Solid lines: plastic phase; dashed lines: liquid phase. (b) Orientational correlation functions for liquid CCl$_4$. (c) Intensities normalized to the asymptotic values of the individual distributions.

We can refine our calculations if we distinguish the bromine and the chlorine atoms. In this way 28 subgroups derive. We concern only the 2:2, the 1:3 and the 3:3 orientations, which can be found in figure 5 for the plastic and the liquid phases. Schematic representations of the most frequently subgroup can be seen in figure 3.c.

Regarding the subgroup level of the 2:2 orientation, the Br,Cl-Br,Cl arrangement is the most frequent edge-to-edge type (2:2) arrangement at the whole distance range. Furthermore, the intensities of the Br,Br-Br,Cl and the Br,Cl-Cl,Cl subgroups exceed the 10%. It also has to be noted that the two phases show almost indistinguishable features in point of the occurrences of the subgroups of the 2:2 and the 1:3 orientations

Subgroups of the 3:3 (face-to-face) orientations, shown in figure 6.c inform us about mutual orientations of the nearest molecules. It is obvious that the most populous arrangement is the Br,Br,Cl-Cl,Cl,Br in the plastic crystalline phase. For liquid phase the Br,Br,Cl-Cl,Cl,Br and the Cl,Cl,Br-Cl,Cl,Br subgroups share the leader role. We have to note that this is the most visible differences, which we found between the two phases.

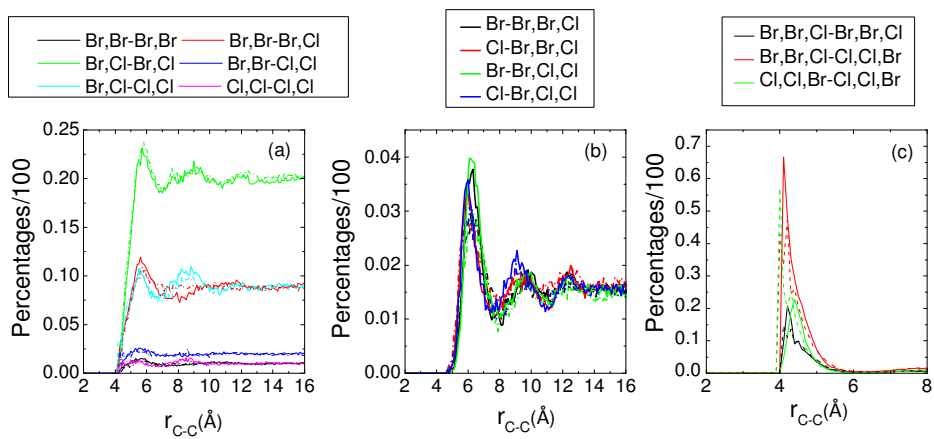

**Figure 5**. (a) The subgroups of the 2:2 orientational correlations. (b) The subgroups of the 1:3 orientational correlations. (c) The subgroups of the 3:3 orientational correlations. Solid lines: plastic phase; dashed lines: liquid phase.

5.3 Dipole-dipole correlations

Although there are many advantages of orientational correlation calculations introduced above, such as the easy visualization, the adaptability for every system with tetrahedral molecules and the unambiguous classification, there is a weak point of this approach: dipole-dipole correlations remain hidden – even in systems of dipolar molecules. The simplest way to gain this type of information is to calculate the angle confined between the two dipole vectors, as a function of carbon-carbon distances [32].

Focusing on the plastic phase, first we have to emphasize that due to the relatively small (around 0.2D) dipole moment of the $CBr_2Cl_2$ molecule, the intensities of the spots in figure 6.a are not high, showing that the dipole effects are not very strong. Still, distinct spots appear at cosine values of -1 and 1, reflecting the parallel and antiparallel arrangements of dipole axes. These spots, that are present also for the liquid state, do not show up beyond 4.5 Å. (It is worth noting that, as a consequence of the $C_{2v}$ symmetry, the dipole vector does not coincide with the carbon-halogen bond, which is a difference from the $CBrCl_3$ molecule.)

Another characteristic angle between dipole vectors is 90°; the corresponding spots (around the cosine value of roughly 0) are confined to a smaller region of carbon-carbon distances. These angles play smaller roles in the liquid phase. As a general remark, the effects of the molecular dipole moments are limited to correlations between neighbours.

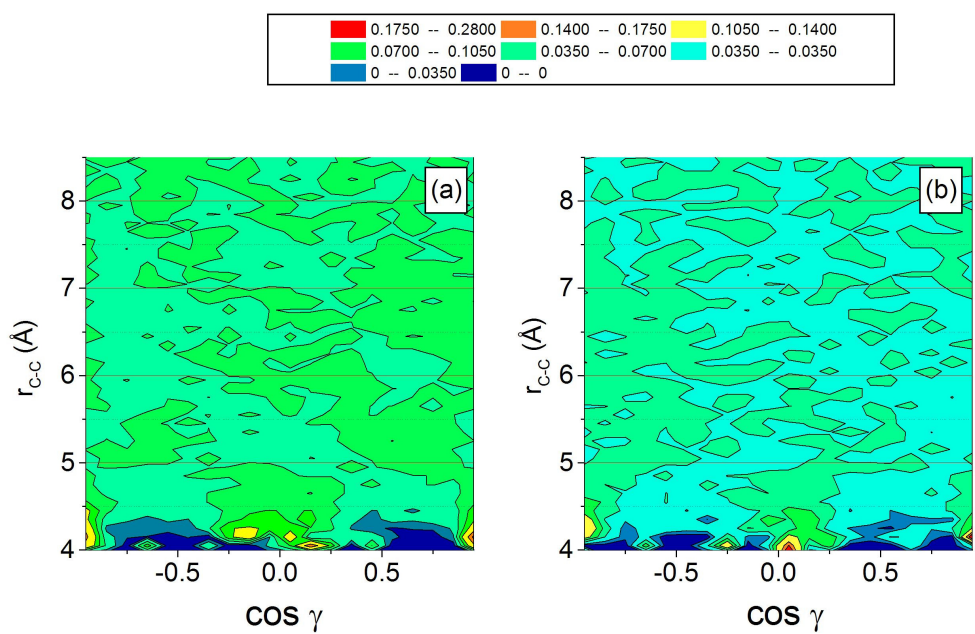

**Figure 6.** Dipole axes correlations. (a) Plastic phase of $CBr_2Cl_2$; (b) Liquid phase of $CBr_2Cl_2$.

6. Conclusions

The atomic-level structure of the plastic phase of $CBr_2Cl_2$ has been studied by neutron diffraction and subsequent Reverse Monte Carlo modeling of the data. A comparison is provided with the corresponding liquid phase at all levels of the structure (total scattering structure factors, partial radial distribution functions, orientational and dipole correlations). Since this compound is the last piece of the extensively studied series of $CBr_nCl_{4-n}$ (n=0, 1, 2, 4), our observations were accompanied by comprehensive remarks.

The plastic and the liquid phases of $CBr_2Cl_2$ show significant similarities. At the touching distances, at and below about 4.5 Å, the molecules prefer the 3:3 (face-to-face) arrangements. Interestingly, in plastic crystalline phase Br,Br,Cl-Cl,Cl,Br arrangements are the most frequent while in the liquid phase the Cl,Cl,Br-Cl,Cl,Br subgroups are also dominant. This latter statement gives the largest differences between the two phases. Increasing the centre-centre distances the Br,Cl-Br,Cl orientations become important.

The results reveal that the relative orientation of neighbouring molecules considerably depends on the steric effect. The influence of the molecular dipole is not as strong as that of the size and shape of the molecules. Our observations go well along with the ones made previously for the $CBr_nCl_{4-n}$ (n=0, 1, 2, 4) series.

Acknowledgment
Financial support was provided by the Hungarian Basic Research Fund (OTKA) Grant No. 083529 and by the Spanish Ministry of Science and Innovation (Grant No. FIS2011-24439), and the Catalan Government (Grant No. 2009SGR-1251). LT is grateful to Drs. G. Bortel and G. Oszlányi (Wigner RCP) for extended discussions.